\documentclass[graybox]{svmult}

\usepackage{helvet}         % selects Helvetica as sans-serif font
\usepackage{courier}        % selects Courier as typewriter font
\usepackage{type1cm}        % activate if the above 3 fonts are
 
\usepackage{cite}
\usepackage{makeidx}         % allows index generation
\usepackage{graphicx}        % standard LaTeX graphics tool
\usepackage{multicol}        % used for the two-column index
\usepackage[bottom]{footmisc}% places footnotes at page bottom

\newcommand\be{\begin{equation}}
\newcommand\ee{\end{equation}}
\newcommand{\bea}{\begin{eqnarray}}
\newcommand{\eea}{\end{eqnarray}}

\newcommand{\nn}{\nonumber}
\newcommand{\pd}{\partial}

\begin{document}

\title*{Primordial Black Hole Generation in a Two-field Inflationary Model}
\titlerunning{PBH Generation in a Two-field Model}
\author{Lilia Anguelova}
\institute{Lilia Anguelova \at Institute for Nuclear Research and Nuclear Energy, Bulgarian Academy of Sciences, Tsarigradsko Chaussee 72, Sofia 1784, Bulgaria,\\ \email{anguelova@inrne.bas.bg}}
\maketitle

\abstract{We summarize our work on the generation of primordial black holes in a type of two-field inflationary models. The key ingredient is a sharp turn of the background trajectory in field space. We show that certain classes of solutions to the equations of motion exhibit precisely this kind of behavior. Among them we find solutions, which describe a transition between an ultra-slow roll and a slow roll phases of inflation.}

\section{Introduction}
\label{sec:1}

Large perturbations during cosmological inflation can seed the formation of Primordial Black Holes (PBHs). If produced with enough abundance, the latter could constitute a significant component of dark matter. This possibility has attracted a lot of attention recently, due to the observation of gravitational waves sourced by binary black hole mergers. The reason is that analysis of the observational data contains indications of primordial origin for a fraction of these black holes \cite{GBSM,FBDLNWBPRV,BC}.

A novel mechanism for the formation of PBHs in multi-field inflation was proposed in \cite{PSZ,FRPRW}. Cosmological models with multiple scalars are of particular interest in view of recent theoretical developments \cite{GK,OPSV,AP,BPR}. An essential difference from the single-field case is that the field-space trajectories of their background solutions can deviate from geodesics. The basic idea of \cite{PSZ,FRPRW} is that a brief period of such a strongly non-geodesic motion can induce a large enhancement of the power spectrum of the curvature perturbation, thus triggering PBH generation. Studying a type of two-field models, we showed in \cite{LA} that there are actual solutions of the background equations of motion, which behave in precisely this manner.  

A key role in the considerations of \cite{LA} was played by a class of exact solutions found in \cite{ABL}. The scalar field space in that case is the Poincar\'e disk, while the scalar potential is determined by a certain hidden symmetry. We review the investigation of \cite{LA}, showing that the field-space trajectories of these solutions exhibit exactly the behavior needed for PBH generation. We also discuss how to improve the problematic behavior of the corresponding Hubble $\eta$-parameter via a suitable symmetry-breaking modification. The resulting modified solutions preserve the PBH-generating properties of the hidden symmetry ones, while describing a smooth transition between an ultra-slow roll and a slow roll inflationary phases.

\section{Two-field inflationary models}
\label{sec:2}

We will study inflationary models arising from two scalar fields $\phi^I (x^{\mu})$ minimally coupled to Einstein gravity. The action describing this system is:
\be \label{Action_gen}
S = \int d^4x \sqrt{-\det g} \left[ \frac{R}{2} - \frac{1}{2} G_{IJ} \pd_{\mu} \phi^I \pd^{\mu} \phi^J - V (\{ \phi^I \}) \right] \,\,\, ,
\ee
where $g_{\mu \nu}$ is the spacetime metric with $\mu,\nu = 0,1,2,3$ and $G_{IJ}$ is the scalar field-space metric with $I,J = 1,2$. As usual in cosmology, we will assume that the background spacetime metric and scalars have the form:
\be \label{metric_g}
ds^2_g = -dt^2 + a^2(t) d\vec{x}^2 \qquad , \qquad \phi^I = \phi^I_0 (t) \quad ,
\ee 
where $a(t)$ is the scale factor. Recall that the Hubble parameter is given by $H(t) = \dot{a}/a$\,, where $\dot{}\equiv \pd_t$\,.

\subsection{Important characteristics}

To define a number of important quantities, characterizing any inflationary model, let us introduce an orthonormal basis of tangent and normal vectors to a field-space trajectory $(\phi^1_0(t),\phi^2_0(t))$\,:
\be
T^I = \frac{\dot{\phi}^I_0}{\dot{\phi}_0} \quad , \quad N_I = (\det G)^{1/2} \epsilon_{IJ} T^J \quad , \quad \dot{\phi}_0^2 = G_{IJ} \dot{\phi}^I_0 \dot{\phi}^J_0 \,\,\, .
\ee
In terms of this basis, the deviation from a geodesic is measured by the quantity \cite{AAGP}:
\be \label{Om_1}
\Omega = - N_I D_t T^I \,\,\, ,
\ee
where $D_t T^I = \dot{\phi}^J_0 \nabla_J T^I$. The function $\Omega (t)$ is called the turning rate of a trajectory. On solutions of the equations of motion, which follow from (\ref{Action_gen}) with (\ref{metric_g}) substituted, the expression (\ref{Om_1}) can be rewritten as:
\be \label{Om_2}
\Omega = \frac{N_I V^I}{\dot{\phi}_0} \,\,\,\, .
\ee

Another set of important characteristics is given by the slow roll parameters, defined in the following manner \cite{CAP}:
\be \label{SR_par}
\varepsilon = - \frac{\dot{H}}{H^2} \qquad , \qquad \eta^I = - \frac{1}{H \dot{\phi}_0} D_t \dot{\phi}_0^I \,\,\,\, .
\ee
Expanding $\eta^I$ in the above basis, we have:
\be
\eta^I = \eta_{\parallel} T^I + \eta_{\perp} N^I \,\,\, ,
\ee
where:
\be \label{eta_PP}
\eta_{\parallel} = - \frac{\ddot{\phi}_0}{H \dot{\phi}_0} \qquad {\rm and} \qquad \eta_{\perp} = \frac{\Omega}{H} \,\,\,\, .
\ee
The phenomenologically-motivated slow roll conditions in the present context are $\varepsilon <\!\!< 1$ and $|\eta_{\parallel}|<\!\!<1$\,. On the other hand, there is no restriction on the dimensionless turning rate $\eta_{\perp}$\,. In fact, we will see shortly that the regime of interest for PBH generation is characterized by $\eta_{\perp}^2 >\!\!> 1$\,.

%\subsection{Perturbations}

To explain the physical mechanism that can seed the formation of primordial black holes, we need to consider perturbations around the homogeneous background (\ref{metric_g}). In comoving gauge, the fields decompose as:
\bea \label{phi_g_decomp}
\phi^I (t, \vec{x}) &=& \phi^I_0 (t) + \delta \phi_{\perp} N^I \,\,\, , \nn \\
g_{ij} (t, \vec{x}) &=& a^2(t) \left[ ( 1 + 2 \zeta ) \delta_{ij} + h_{ij} \right] \,\,\, ,
\eea
where $\delta \phi_{\perp} (t, \vec{x})$ is the entropic perturbation, $\zeta = \zeta (t,\vec{x})$ is the curvature one and $h_{ij} (t, \vec{x})$ are tensor fluctuations with $i,j = 1,2,3$ being spatial indices. Substituting (\ref{phi_g_decomp}) in (\ref{Action_gen}), one can derive an effective action for the perturbations. The key ingredients in that action are an interaction term between $\zeta$ and the entropic perturbation, as well as a mass term for $\delta \phi_{\perp}$ of the form (see, for instance, \cite{AAGP}): 
\be \label{m2_eff_gen}
m_s^2 = N^I N^J V_{;IJ} - \Omega^2 + \varepsilon H^2 \mathcal{R} \,\,\, ,
\ee
where $V_{;IJ} = \pd_I \pd_J V - \Gamma^K_{IJ} V_K$ and $\mathcal{R}$ is the Ricci scalar of the field-space metric $G_{IJ}$. 

The interaction term, whose strength depends on $\eta_{\perp}$\,, implies that $\delta \phi_{\perp}$ can affect the evolution of $\zeta$\,, and thus of the density fluctuations in the Early Universe. In particular, the amplitude of the curvature perturbation can become significantly enhanced for large enough turning rate \cite{PSZ,FRPRW}. The latter, however, can induce a negative entropic mass in view of (\ref{m2_eff_gen}). Thus, a brief tachyonic instability of the entropic perturbation can signify the formation of primordial black holes. Our goal will be to show that there are actual solutions to the background equations of motion, which lead to precisely this kind of behavior for $\eta_{\perp} (t)$ and $m_s^2 (t)$\,.

\subsection{Rotationally invariant field spaces}

Let us now focus on rotationally invariant scalar field spaces. Then we can write the metric $G_{IJ}$ as:
\be \label{Gmetric}
ds^2_{G} = d\varphi^2 + f(\varphi) d\theta^2 \,\,\, ,
\ee
where we have denoted:
\be \label{Backgr_id}
\phi^1_0 (t) \equiv \varphi (t) \qquad {\rm and} \qquad \phi^2_0 (t) \equiv \theta (t) \,\,\, . 
\ee

Using (\ref{Gmetric})-(\ref{Backgr_id}) together with (\ref{metric_g}), one finds from (\ref{Action_gen}) the following equations of motion for the background:
\be \label{ScalarEoMs}
\ddot{\varphi} - \frac{f'}{2} \dot{\theta}^2 + 3 H \dot{\varphi} + \pd_{\varphi} V = 0 \quad , \quad \ddot{\theta} + \frac{f'}{f} \dot{\varphi} \dot{\theta} + 3 H \dot{\theta} + \frac{1}{f} \pd_{\theta} V = 0  \quad ,
\ee
\be \label{EinstEq}
\dot{\varphi}^2 + f \dot{\theta}^2 = - 2 \dot{H} \quad , \quad 3H^2 + \dot{H} = V \quad .
\ee

We will be interested specifically in the case with $\pd_{\theta} V = 0$\,. In that case, (\ref{Om_2}) gives for the turning rate \cite{LA}:
\be \label{Om_PD}
\Omega = \frac{\sqrt{f}}{\left( \dot{\varphi}^2 + f \dot{\theta}^2 \right)} \,\dot{\theta} \,\pd_{\varphi} V \,\,\, ,
\ee
In addition, the effective entropic mass (\ref{m2_eff_gen}) acquires the form \cite{LA}:
\be 
\label{m_s_f}
m_s^2 = M^2_V - \Omega^2 + \varepsilon H^2 \mathcal{R} \,\,\, ,
\ee
where:
\be \label{MV}
M^2_V \equiv \frac{ f \dot{\theta}^2 \pd_{\varphi}^2 V + \frac{f'}{2 f} \dot{\varphi}^2 \pd_{\varphi} V }{ ( \dot{\varphi}^2 + f \dot{\theta}^2 ) } \,\,\, .
\ee

We should note that $\Omega (t)$ can be a non-trivial function, i.e. the background trajectories in field space can be genuinely non-geodesic, even though the potential does not depend on one of the two scalars, as will become clear shortly; see also \cite{LA} and references therein.

\section{A class of exact solutions}
\label{sec:3}

Now we will show that a class of exact solutions to (\ref{ScalarEoMs})-(\ref{EinstEq}), obtained in \cite{ABL},  leads to a brief period with large turning rate, as well as tachyonic entropic mass, as needed for PBH generation. These solutions arise from the following choices of the functions $f$ and $V$\,:
\be \label{f_D}
f (\varphi) \,= \,\frac{8}{3} \,\sinh^2 \!\left( \sqrt{\frac{3}{8}} \,\varphi \right) \,\,\,\, ,
\ee
\be \label{Pot}
V (\varphi , \theta) \,= \,V_0 \,\cosh^2 \!\left( \sqrt{\frac{3}{8}} \,\varphi \right) \,\,\,\, .
\ee
Note that (\ref{f_D}) is equivalent with taking the field-space metric (\ref{Gmetric}) to be that of the Poincar\'e disk (with fixed Gaussian curvature). Then (\ref{Pot}) is exactly the form of the potential required by the hidden symmetry of \cite{ABL}. 

For the above choices of potential and field space, one can solve (\ref{ScalarEoMs}) by:
\bea \label{Sols_asc}
a (t) &=& \left[ u^2 - \left( v^2 + w^2 \right) \right]^{1/3} \,\,\,\, , \nn \\
\varphi (t) &=& \sqrt{\frac{8}{3}} \,{\rm arccoth} \!\left( \sqrt{\frac{u^2}{v^2 + w^2}} \,\,\right) \,\,\,\, , \nn \\
\theta (t) &=& {\rm arccot} \!\left( \frac{v}{w} \right) \,\,\,\, ,
\eea
where $u$, $v$ and $w$ are the following functions:
\bea \label{Sols_uvw}
u (t) &=& C^u_1 \sinh \!\left( \kappa \,t \right) + C^u_0 \cosh \!\left( \kappa \,t \right) \,\,\, , \,\,\, \kappa \equiv \frac{1}{2} \sqrt{3 V_0} \quad \,, \nn \\
v (t) &=& C_1^v \,t + C_0^v \qquad {\rm and} \qquad w (t) = C_1^w \,t + C_0^w \quad \,,
\eea
with $C^{u,v,w}_{0,1} = const$\,. The expressions (\ref{Sols_asc})-(\ref{Sols_uvw}) solve (\ref{EinstEq}) as well, if the following relation between the integration constants is satisfied:
\be \label{Constr_s}
(C_1^v)^2 + (C_1^w)^2 \,= \,\kappa^2 \left[ (C_1^u)^2 - (C_0^u)^2 \right] \,\,\, .
\ee

Substituting (\ref{Sols_asc}), together with (\ref{f_D})-(\ref{Pot}), inside (\ref{Om_PD}) and (\ref{MV}) gives:
\be \label{Om_uvw}
\Omega \,= \frac{3V_0}{4} \,\frac{u \,(v \dot{w} - \dot{v} w) \,\sqrt{u^2-w^2-v^2}}{\left[ (v \dot{u} - \dot{v} u)^2 + (w \dot{u} - \dot{w} u)^2 - (v \dot{w}-\dot{v} w)^2 \right]}
\ee
and
\be \label{m2_uvw}
M_V^2 = \frac{3V_0}{4} \,\frac{ \left\{ u^2 \left[ (v \dot{u} - \dot{v} u)^2 + (w \dot{u} - \dot{w} u)^2 \right] - (v^2 + w^2) (v \dot{w}-\dot{v} w)^2 \right\}}{(u^2 - v^2 - w^2) \left[ (v \dot{u} - \dot{v} u)^2 + (w \dot{u} - \dot{w} u)^2 - (v \dot{w}-\dot{v} w)^2 \right]} \,\,\, ,
\ee
respectively. Analyzing these expressions directly, upon substitution of (\ref{Sols_uvw}), is rather daunting. So \cite{LA} used a combination of analytical and numerical means to understand their behavior.

For that purpose, it is very useful to introduce the canonical radial variable on the Poincar\'e disk, $\rho$, which runs in the range $0 \le \rho < 1$ and is related to the field $\varphi$ via: 
\be \label{rho_phi}
\rho = \tanh \!\left( \frac{1}{8} \sqrt{6} \,\varphi \right) \,\, .
\ee
Using (\ref{Sols_asc}) in (\ref{rho_phi}), one finds:
\be \label{rho_uvw}
\rho (t) = \frac{\sqrt{v^2 + w^2}}{\sqrt{u^2 - v^2 - w^2} + \sqrt{u^2}} \,\,\,\, ,
\ee
which implies that the extremum condition $\dot{\rho} (t) = 0$ can be written as:
\be \label{rho_extr_eq}
( v^2 + w^2 ) \,\dot{u} - \left( v \dot{v} + w \dot{w} \right) u = 0 \,\,\, .
\ee

\begin{figure}[t]
%\begin{figure}[h!]
\begin{center}
\hspace*{-0.3cm}
\includegraphics[scale=0.25]{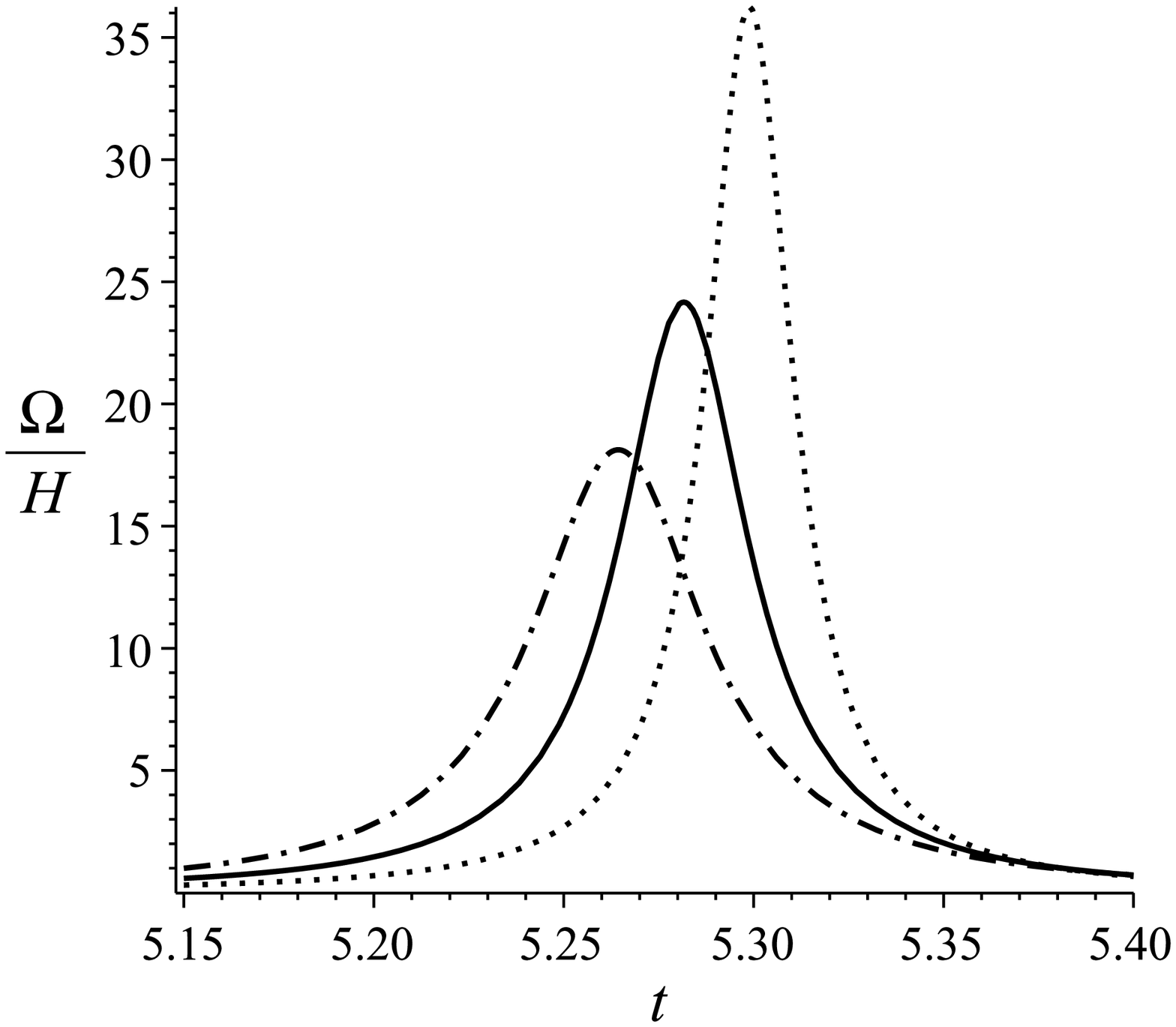}
\hspace*{0.3cm}
\includegraphics[scale=0.272]{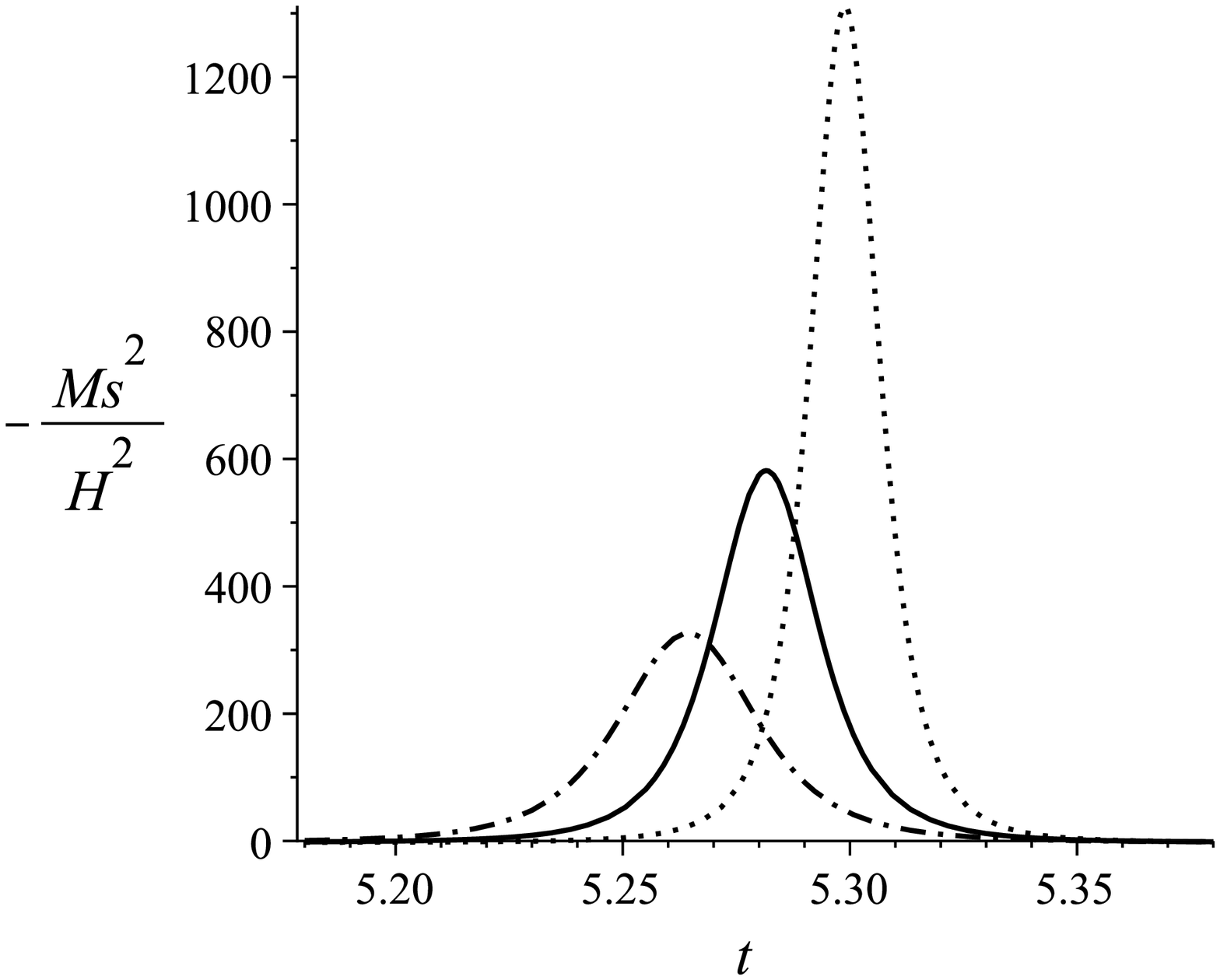}
\end{center}
\vspace{-0.2cm}
\caption{{\small Three examples of the functions $\Omega / H$ and $-m_s^2 / H^2$\,, obtained by taking $\kappa = 3$, $C_0^u = 6$, $C_0^v = 1$, $C_1^v = -\frac{1}{5}$, $C_1^w = \frac{1}{2}$ and $C_1^u$ - the positive root of (\ref{Constr_s}), as well as $C_0^w = - 2.46$ (dash-dotted curve), $C_0^w = -2.47$ (solid curve) and $C_0^w = -2.48$ (dotted curve).}}
\label{Om_Ms}
\vspace{0.1cm}
\end{figure}
Analyzing (\ref{rho_extr_eq}), with (\ref{Sols_uvw}) substituted, \cite{LA} showed that the function $\rho (t)$ can have at most two local extrema. In the cases with a single extremum, the latter is always a maximum. In the cases with two extrema, the one occurring at an earlier time is a local minimum, whereas the one occurring later is a local maximum. In any case, only the local maximum corresponds to a turn of the trajectory. Furthermore, the single turn induces a peak of $\Omega (t)$, as well as a corresponding peak of $-m_s^2 (t)$\,. We have illustrated the behavior of these functions on Figure \ref{Om_Ms}. The examples, plotted there, have been chosen for convenience.\footnote{In particular, in all three examples \,$\varepsilon <\!\!< 1$ (in fact, $\varepsilon (t)|_{t_{peak}} \sim 10^{-19}$) and $H=2$ to a great degree of accuracy, implying that the number of e-folds is $N = 2 t$\,.} It should be stressed, though, that the shape of the functions is always the same, regardless of the values of the integration constants. However, as demonstrated in \cite{LA}, both the position and the height of the peak can be varied at will by choosing suitably the values of the constants. In particular, one can easily achieve numerically that $|\eta_{\perp}|_{t_{peak}} \approx 23$\,, which is necessary for triggering PBH generation according to the benchmark cases of \cite{FRPRW}. Note that, interestingly, the rapid turn occurs during a single e-fold.

The above results show that, in principle, the exact solutions (\ref{Sols_asc})-(\ref{Constr_s}) are suitable for describing the generation of perturbations, large enough to seed PBH formation. Unfortunately, however, it turns out that the corresponding $\eta_{\parallel}$-parameter violates the slow roll approximation \cite{LA}. Since $\eta_{\parallel}$ equals the Hubble slow roll parameter $\eta_H \equiv - \frac{\ddot{H}}{2 H \dot{H}}$ on solutions of the equations of motion, this means that the inflationary regime under consideration is not viable phenomenologically. In the next Section, we will show how one can remedy the behavior of $\eta_{\parallel}$\,, while preserving the desired for PBH-generation properties of the solutions.

\section{Modified solution and PBH generation}
\label{sec:4}

Our goal now is to find modified solutions of the equations of motion, which preserve the desirable behavior of $\eta_{\perp} (t)$\,, while improving that of $\eta_{\parallel} (t)$\,. For that purpose, let us consider the following Ansatze:
\be \label{f_q}
f (\varphi) \,= \,\frac{1}{q^2} \,\sinh^2 (q \varphi) \,\,\, ,
\ee
\be \label{V_q}
V (\varphi , \theta) \,= \,V_0 \,\cosh^{6p} (q \varphi) \,\,\, ,
\ee
where $p,q=const$ and $p>0$\,. Thus, the field space metric is still that of the Poincar\'e disk, although for arbitrary $q$ the hidden symmetry of \cite{ABL} is not preserved. Let us also introduce new variables $\tilde{u}, \tilde{v}, \tilde{w}$ via the Ansatz:
\bea \label{uvw_mod}
\tilde{u} &=& a^{\frac{1}{2p}} \,\cosh (q \varphi) \,\,\, , \nn \\
\tilde{v} &=& a^{\frac{1}{2p}} \,\sinh (q \varphi) \,\cos \theta \,\,\, , \nn \\
\tilde{w} &=& a^{\frac{1}{2p}} \,\sinh (q \varphi) \,\sin \theta \,\,\, .
\eea
Notice that, taking \,$p=\frac{1}{3}$ \,and \,$q=\sqrt{\frac{3}{8}}$ \,inside (\ref{f_q})-(\ref{uvw_mod}), one recovers precisely the expressions (\ref{f_D})-(\ref{Sols_asc}), relevant for the exact solutions with hidden symmetry that we considered above. In addition, one can show that the equations of motion, resulting from (\ref{f_q})-(\ref{uvw_mod}), simplify significantly for:
\be \label{qp}
q = \frac{1}{\sqrt{24}} \,\frac{1}{p} \,\,\, .
\ee
So we will assume (\ref{qp}) from now on, as well.

In \cite{LA} it was argued that a phenomenologically preferable regime, ensuring that $\varepsilon<\!\!<1$ at early times, is obtained in the large-$\tilde{u}$ limit:
\be \label{Approx_reg}
|\tilde{u}|,|\dot{\tilde{u}}| \,>\!\!> \,|\tilde{v}|, |\tilde{w}|, |\dot{\tilde{v}}|, |\dot{\tilde{w}}| \quad .
\ee
In this regime, the equations of motion, that follow from (\ref{f_q})-(\ref{qp}), acquire the form:
\bea
24 \,p \,\tilde{u} \,\ddot{\tilde{u}} + 24 \,p \,(3p-1) \,\dot{\tilde{u}}^2 - 6 \,V_0 \,\tilde{u}^2 &=& 0 \,\,\, , \label{ut_EoM} \\
\ddot{\tilde{y}} + 2 (3p-1) k_u \dot{\tilde{y}} - (3p-1) k_u^2 \tilde{y} &=& 0 \,\,\, , \label{yt_EoM}
\eea
where $\tilde{y} = \tilde{v}, \tilde{w}$\,. These equations can be solved respectively by:
\be \label{u_mod_sol}
\tilde{u}(t) \,= \,C_u \,e^{k_u t} \quad , \quad k_u = \sqrt{\frac{V_0}{12}} \,\frac{1}{p}
\ee
and
\be \label{vw_mod_sol}
\tilde{v} (t) = C_v e^{k_v t} \quad , \quad \tilde{w} (t) = C_w e^{k_w t} \,\,\, ,
\ee
where
\bea \label{kvkw}
k_v &=& - k_u \left[ (3p-1) + \sqrt{(3p-1)3p} \,\right] \,\,\, , \nn \\ 
k_w &=& - k_u \left[ (3p-1) - \sqrt{(3p-1)3p} \,\right] \,\,\, .
\eea
Note that substituting $p = 1/3$ inside (\ref{ut_EoM})-(\ref{yt_EoM}) leads precisely to the hidden symmetry solutions in (\ref{Sols_uvw}). Also, to have real $k_{v,w}$ in (\ref{kvkw}), we need $p \ge \frac{1}{3}$\,. In fact, we will take \,$p>1$ \,from now on, to ensure a phenomenologically desirable behavior of \,$\eta_{\parallel}$ \,according to the discussion in \cite{LA}.

Now, using (\ref{f_q})-(\ref{V_q}) and (\ref{u_mod_sol})-(\ref{kvkw}), together with the inverse of (\ref{uvw_mod}), one can show that the turning rate $\Omega (t)$ in (\ref{Om_PD}) has a single peak. Furthermore, one can compute analytically the position, height and width of the peak \cite{LA}. Similarly, one can show analytically that the corresponding effective entropic mass $m_s^2 (t)$ in (\ref{m_s_f}) has a transient tachyonic instability \cite{LA}. Specifically, we have that:
\be \label{m2_mod_peak}
m_s^2|_{t=t_{peak}} = (m_V^2 - \Omega^2)|_{t=t_{peak}} = - 3 k_u^2 p \,(3p-2) \,\,\,\, ,
\ee
whereas before and after the peak:
\be \label{m2_mod_as}
m_s^2 \, = \, m_V^2 - \Omega^2 \,\,\, \rightarrow \,\,\, 3 k_u^2 p \qquad \, {\rm as} \, \qquad t \rightarrow 0 \,\,\,\,\, {\rm or} \,\,\,\,\, t \rightarrow \infty \quad .
\ee
Here we have used that $\varepsilon \!<\!\!< \!1$ in the entire range of validity of the new solutions (\ref{u_mod_sol})-(\ref{kvkw}), as well as that the Ricci scalar of the field-space metric is fixed to ${\cal R} = - \frac{1}{12 p^2}$\,.

From the inverse of (\ref{uvw_mod}), one can also find the $\eta_{\parallel}$-parameter of the new solutions \cite{LA}; recall that $\eta_{\parallel} (t) = - \frac{\ddot{H}}{2 H \dot{H}}$\,. At early times, i.e. for $t \approx 0$\,, the full analytical expression reduces to:
\be \label{eta_mod_t0}
\eta_{\parallel} \, \approx \, \frac{(k_u - k_v)}{2pk_u} = \frac{3p+\sqrt{(3p-1)3p}}{2p} \,\,\, .
\ee
Note that this is well-approximated numerically by $\eta_{\parallel} \approx 3$ for any $p > 2$\,. Thus, before the turn one has an ultra-slow roll inflationary phase. On the other hand, at late times, i.e. for large $t$\,, we have:
\be
\eta_{\parallel} \, \approx \, \frac{(k_u - k_w)}{2pk_u} = \frac{3p-\sqrt{(3p-1)3p}}{2p} \,\,\, ,
\ee
which is well-approximated numerically by $\eta_{\parallel} \approx \frac{1}{4p}$ for any $p > 2$\,. Clearly, by choosing a suitably large value of $p$\,, we can ensure that the slow roll approximation $\eta_{\parallel} <\!\!< 1$ is well satisfied after the turn. Hence, the modified solutions of this Section describe a smooth transition between an ultra-slow roll and a slow roll inflationary phases, for any $p$ greater than $4$ or so.\footnote{This is similar to the numerical results of \cite{BHFSSS}, where PBH generation is also triggered by a transition between two phases of inflation. However, in that work the transition is due to a separate potential term for each scalar, driving its own slow-roll expansion phase.}

We should note that the solutions (\ref{u_mod_sol})-(\ref{kvkw}) are approximate, since they were derived in the large-$\tilde{u}$ limit. However, they satisfy (\ref{Approx_reg}) more and more accurately with time. Hence, as discussed in \cite{LA}, one can improve them by considering small corrections, at early times, of the form:
\bea \label{Mod_cor}
\tilde{v}(t) &=& \left( \,C_v + C_v^{(1)} \,t + C_v^{(2)} \,t^2 + ... \,\right) e^{k_v t} \,\,\,\, , \nn \\
\tilde{w}(t) &=& \left( \,C_w + C_w^{(1)} \,t + C_w^{(2)} \,t^2 + ... \,\right) e^{k_w t} \,\,\,\, ,
\eea
where $C_{v,w}^{(1),(2),...} = const$\,. Such subleading corrections in $\tilde{v} (t)$ and $\tilde{w} (t)$ leave the slow roll parameters $\varepsilon (t)$ and $\eta_{\parallel} (t)$ essentially unchanged, since the scale factor $a(t)$ is dominated by $\tilde{u} (t)$\,. However, the function $\dot{\theta} (t)$ is rather sensitive to the corrections in (\ref{Mod_cor}), implying that $\eta_{\perp} (t)$ is as well. Thus, as demonstrated in \cite{LA}, the sharpness of the turn of a trajectory in field space can be significantly affected by the above subleading corrections. In view of this, it is easy to obtain any magnitude of $\eta_{\perp} (t_{peak})$\,, desirable for PBH-generation, by choosing suitably the values of the constants. Interestingly, the corresponding sharp turn can last several e-folds.

Finally, it is worth pointing out that, in the large $\tilde{u}$-regime, one is always near the center of field space. Specifically, our slow roll phase occurs for $\varphi <\!\!< 1$\,. This is in stark contrast with the usual hyperbolic models in the literature, which rely on large field values (even close to the boundary of the Poincar\'e disk, which is at $\varphi \rightarrow \infty$) to achieve slow roll expansion. Thus our models provide a much more reliable effective description of the inflationary period.

\begin{acknowledgement}
I have received partial support from the Bulgarian NSF grants DN 08/3 and KP-06-N38/11.
\end{acknowledgement}
%
%\section*{Appendix}
%\addcontentsline{toc}{section}{Appendix}
%
%

%\input{referenc}
\end{document}